\documentclass[11pt, A4]{article}

\usepackage{graphicx}
\usepackage{epstopdf}
\usepackage{amsmath}
\usepackage{amssymb}
\usepackage{amsfonts}
\usepackage{amsthm}
\usepackage[usenames]{color}
\usepackage{array}
\usepackage{tabularx}
\usepackage{booktabs}           
\usepackage{caption}
\usepackage{multirow}
\usepackage[usenames,dvipsnames,svgnames,table]{xcolor}
\usepackage[utf8]{inputenc}
\usepackage{ucs}
\usepackage{lmodern}
\usepackage{verbatim}
\usepackage{mdframed}
\usepackage{lipsum}
\usepackage{amsmath}
\usepackage{amsfonts}
\usepackage{amssymb}
\usepackage{eso-pic}
\usepackage{graphicx}
\usepackage[final]{pdfpages}
\usepackage{pdfpages}
\usepackage[usenames,dvipsnames,svgnames,table]{xcolor}
\usepackage{cancel}
\numberwithin{equation}{section}

\usepackage[
      colorlinks=true,
      linkcolor=blue,
      urlcolor=blue,
      filecolor=blue,
      citecolor=red,
      pdfstartview=FitV,
      pdftitle={},
      pdfauthor={},
      pdfsubject={},
      pdfkeywords={},
      pdfpagemode=None,
      bookmarksopen=true
]{hyperref}

\usepackage{empheq}
\usepackage{mathtools}
\usepackage{epsfig}

\usepackage{hyperref}

\def\beq{\begin{eqnarray}}
\def\eeq{\end{eqnarray}}

\def\g{\gamma}

\def\be{\begin{equation}}
\def\ee{\end{equation}}
\def\bea{\begin{eqnarray}}
\def\eea{\end{eqnarray}}

\newcommand{\rom}[1]{\mathrm{#1}}

\def\cF{\mathcal{F}}

\def\nn{\nonumber}

\definecolor{X}{rgb}{0,0,1}
\definecolor{Y}{rgb}{1,0,0}
\definecolor{Z}{rgb}{0,51,0}

\newcommand{\del}{\nabla}
\def\bes{\begin{subequations}}
\def\ees{\end{subequations}}

\numberwithin{equation}{section}

\renewcommand{\thefootnote}{\fnsymbol{footnote}}

\begin{document}

\begin{centering}

\thispagestyle{empty}

{\flushright {Preprint IMSc, NISER, CMI 2019}\\[15mm]}

{\LARGE \textsc{Generalised Garfinkle-Vachaspati  Transform \\ {\vskip 0.3cm} With Dilaton}} \\

 \vspace{0.8cm}

{\large 

Subhroneel Chakrabarti$^{1, 3}$, Deepali Mishra$^{2, 3}$, Yogesh K.~Srivastava$^{2, 3}$, \\ and  Amitabh Virmani$^{4}$}
\vspace{0.8cm}

\begin{minipage}{.9\textwidth}\small  \begin{center}
${}^{1}${The Institute of Mathematical Sciences,  CIT Campus, \\
Taramani, 
Chennai, 
Tamil Nadu, India 600 113}\\
  \vspace{0.5cm}
${}^{2}${National Institute of Science Education and Research (NISER), \\ Bhubaneswar, P.O. Jatni, Khurda, Odisha, India 752050}\\
  \vspace{0.5cm}
$^3$Homi Bhabha National Institute, Training School Complex, \\ Anushakti Nagar, Mumbai, India 400085 \\
  \vspace{0.5cm}
$^4$Chennai Mathematical Institute, H1 SIPCOT IT Park, \\ Kelambakkam, Tamil Nadu, India 603103\\
  \vspace{0.5cm}
{\tt subhroneelc@imsc.res.in, deepalimishra, yogeshs@niser.ac.in, \\ avirmani@cmi.ac.in}
\\ $ \, $ \\

\end{center}
\end{minipage}

\end{centering}

\renewcommand{\thefootnote}{\arabic{footnote}}

\begin{abstract} 

The generalised Garfinkle-Vachaspati transform (GGV) introduced in our previous work (arXiv:1808.04981) is a powerful technique to add hair modes to a class of known D1-D5 solutions. The richest and most interesting examples of D1-D5 geometries involve the dilaton and this calls for an extension of the procedure to accommodate for the presence of the dilaton field. In this paper we present such extended version of the GGV transform. We explore this generalisation in ten-dimensions with R-R or NS-NS 2-form field and relate the two set-ups via S-duality.
 In the context of the D1-D5 system, this generalisation allows us to add travelling wave deformations on solutions beyond minimal six-dimensional supergravity lifted to IIB supergravity.  We work out travelling wave deformations involving the torus directions on a class of supersymmetric D1-D5-P geometries. 
We also explore applications of our technique to the F1-P system.    
\end{abstract}

\newpage
\tableofcontents

\setcounter{equation}{0}

\section{Introduction}

String theory has had many successes with black holes, including precision counting of  microscopic degeneracies for certain supersymmetric brane systems relevant to black hole physics~\cite{Sen:1995in, Strominger:1996sh, Sen:2007qy, Mandal:2010cj, Dabholkar:2012zz}. In these discussions a concept of black hole ``hair'' plays an important role~\cite{Banerjee:2009uk, Jatkar:2009yd, Mathur:2018tib}. The term black hole hair refers to normalisable deformations of the black hole solution with support outside the horizon. It is often the case that the microscopic degeneracies formulae count both the degeneracy of the black hole and the contributions due to hair modes.

An example is  the 4D-5D lift relating BMPV black hole to a four dimensional black hole. These two set-ups have identical near horizon geometries, but different microscopic degeneracies of the corresponding brane systems~\cite{David:2006yn}. This difference in the microscopic degeneracies can be accounted for by the hair modes~\cite{Banerjee:2009uk, Jatkar:2009yd}.

In construction of such hair modes, the Garfinkle-Vachaspati (GV) transform~\cite{Garfinkle:1990jq} is of fundamental importance.  The technique allows to add travelling wave deformations (hair modes) on the above mentioned black holes.  It goes as follows: given a spacetime  with metric $g_{\mu \nu }$ admitting a null, Killing, and hypersurface orthogonal vector field $k^\mu$, i.e., 
\begin{align}
  k^\mu k_\mu &= 0, &
    \del_{(\mu} k_{\nu)}  &=0, &   \del_{[\mu} k_{\nu]}  &= k_{[\mu}\del_{\nu]} S,\label{ordinary_GV1}
\end{align}
for some scalar function $S$,  one can construct a new exact solution
of the equations of motion as,
\begin{equation}
g'_{\mu \nu}= g_{\mu\nu} + e^{-S} \, \chi \, k_{\mu }k_{ \nu},\label{ordinary_GV2}
\end{equation}
where  $\chi$ satisfies massless wave equation with respect to the background metric,
\begin{align}
\Box \chi &= 0, & k^{\mu}\partial_{\mu}\chi &=0. \label{ordinary_GV3}
\end{align}
Matter fields, if present, do not transform as long as some transversality conditions on form-fields are satisfied. For further details we refer the reader to~\cite{Kaloper:1996hr}. The technique has been applied in varied contexts, see e.g.,~\cite{Jatkar:2009yd, Kaloper:1996hr,Dabholkar:1995nc, Horowitz:1996th, Banados:1999tw, Hubeny:2003ug,  Balasubramanian:2010ys}.
Not only the techniques that add  travelling waves allow us to construct hair modes on certain class of black holes, they are of crucial importance in the construction of black hole microstates in the context of the fuzzball paradigm~\cite{Mathur:2005zp, Bena:2007kg, Skenderis:2008qn, Chowdhury:2010ct}. 
For example, travelling wave solutions describing  F1-P bound states -- often referred to as  Dabholkar-Harvey bound states~\cite{Dabholkar:1995nc, Callan:1995hn} -- can be dualised into two charge D1-D5 bound states~\cite{Lunin:2001jy, Mathur:2005zp}.  A natural question is: can one add further hair on D1-D5 bound states using the Garfinkle-Vachaspati transform?

Unfortunately, the technique does not find a useful application in the context of the D1-D5 bound states. For these geometries  the null Killing vector fails to be  hypersurface orthogonal.  This roadblock led authors to explore alternative perturbative techniques to construct hair modes on the known D1-D5 geometries \cite{Mathur:2011gz, Mathur:2012tj}. In \cite{Lunin:2012gp} it was speculated that a generalised Garfinkle-Vachaspati transform might exist that does not require the hypersurface orthogonality property of the null Killing vector.

Recently, we proposed a generalised Garfinkle-Vachaspati (GGV) transform~\cite{GGV} that finds applications in the context of D1-D5 geometries too.\footnote{The name generalised Garfinkle-Vachaspati (GGV)  transform is somewhat confusing as it seems to suggest that there is a limit of the GGV transform  to the original  GV transform, but this is not the case. In a subject that is so saturated with various ``generalised" mathematical objects perhaps a more specific name for the transformations could be more appropriate. Though, we have chosen to stick to this name as typically in the context of the D1-D5 geometries, the original  GV transform does not find applications but these generalisations do.} It allows us to add travelling wave deformations to certain solutions of minimal six-dimensional supergravity embedded in ten-dimensional type IIB theory via trivially adding the four-torus directions.  The travelling wave deformations involve the four-torus directions.  Not only the metric but also the Ramond-Ramond 2-form field supporting the solution transforms in a specific way.

A priori it is not at all obvious if the generalised Garfinkle-Vachaspati transform (GGV) can be extended to other set-ups. The most pertinent extension is the inclusion of the dilaton. This is because,  (i) most of the interesting and rich examples of D1-D5 geometries involve the dilaton, (ii) the presence of the dilaton allows one to convolute the GGV technique with S-duality.  In this paper we present a generalisation of the generalised Garfinkle-Vachaspati transform with dilaton. As detailed in appendix \ref{app:EOM_analysis}, it is quite non-trivial that the technique finds a generalisation with dilaton.

The rest of the paper is organised as follows. In section \ref{sec:GGV} we postulate our generalised Garfinkle-Vachaspati transform (GGV) with dilaton.  Two  set-ups related by S-duality, namely, 
 ten-dimensional type IIB Ramond-Ramond (R-R) sector with dilaton and
 the ten-dimensional Neveu-Schwarz (NS-NS) sector with dilaton, are addressed in  sections  
  \ref{sec:GGV_RR} and
 \ref{sec:GGV_NS} respectively.
In section \ref{sec:D1-D5} we work out travelling-wave deformation involving the torus directions on a class of supersymmetric D1-D5-P orbifold geometries. The deformed solutions are given in terms of solutions of a (non-minimally coupled)  scalar field on the background geometry.
When the background contains a large AdS region, the deformed states are identified in the D1-D5 CFT as an action of a U(1) current on the undeformed state. In section \ref{sec:F1-P}   application of the GGV technique to the F1-P system is discussed. In section \ref{sec:disc} we close with a brief discussion of possible future directions. Technical details involved in establishing the GGV technique are presented in appendix \ref{app:EOM_analysis}.

\section{Generalised Garfinkle-Vachaspati transform with dilaton}
\label{sec:GGV}
Generalised Garfinkle-Vachaspati (GGV) transform was introduced in \cite{GGV} as a novel solution generating technique in IIB theory and in various other duality frames. The technique allows to add wave like deformations to certain solutions of minimal six-dimensional supergravity trivially lifted to IIB theory along the four-torus directions. In this paper we explore a generalisation of the GGV transform where the six-dimensional solutions can have a non-trivial dilaton profile. There are two closely related set-ups for which this generalisation is  explored in this paper: (i) ten-dimensional type IIB Ramond-Ramond (RR) sector with dilaton and the two-form RR field, and (ii) ten-dimensional Neveu-Schwarz (NS-NS) sector with dilaton.

\subsection{Transform for the type IIB R-R sector}
\label{sec:GGV_RR}
The non-zero dilaton brings in several new elements. From the six-dimensional perspective, in general, we can no longer truncate to minimal supergravity. The simplest set-up in six dimensions that allows for the dilaton is minimal supergravity coupled to one self-dual tensor multiplet. 
 The six-dimensional action is, see, e.g.~\cite{Duff},
 \be
 S_6 = \frac{1}{16\pi G_6} \int d^6x \sqrt{-g} \left[ R - (d\phi)^2  - \frac{1}{12} e^{2 \phi} F_{\mu \nu \rho}F^{\mu \nu \rho}\right]. \label{action_6d_main_text}
 \ee
 where $F^{(3)}=d C^{(2)}$. This is the 6d theory we will work with exclusively.  For ten-dimensional fields we follow Polchinski's conventions \cite{Pol1, Pol2}. The
ten-dimensional IIB string frame action with R-R 2-form field is 
\be
S_\rom{RR}=\frac{1}{16 \pi G_{10}}\int d^{10}x\sqrt{-g}\left[e^{-2\Phi}[R+4(d\Phi)^2]-\frac{1}{12}F_{\mu \nu \rho}F^{\mu \nu \rho}\right].
\ee

 The embedding of interest  of six-dimensional fields in ten-dimensions is
\be
ds^2_{\rom{(S)}}=ds_{6}^2+e^{\phi} ds_4^2,\label{metric_string_10d_main_text}
\ee
where $ds^2_{\rom{(S)}}$ is the ten-dimensional string frame metric, $ds^2_4 = \sum_{i=1}^{4} dz^i dz^i$ is the flat torus metric, $\phi$ is the six-dimensional dilaton. The ten-dimensional dilaton is same as the six-dimensional dilaton 
\be
\Phi =\phi,
\ee 
and the ten-dimensional 2-form R-R field is also same as the six-dimensional 2-form field with zero components in the four torus directions.

The spacelike Killing vectors provided by the torus directions,
\be
l_{(i)}= l_{(i)}^\mu \partial_\mu =\partial_{z^{i}},
\ee 
are normalised as $l^\mu l_\mu =e^{\phi}$. These vectors are not  covariantly constant, unlike in the analysis of \cite{GGV}.  Let $k^\mu$ be a null Killing vector of the six-dimensional metric $ds^2_6$, with the property that the dilaton is compatible with the null Killing symmetry\footnote{To avoid notational clutter we have not introduced separate indices that range over six-dimensional spacetime. Most of the equations we write are in ten-dimensions. It should be clear from the context when the indices run over six dimensions.}
\be
k^\mu\partial_\mu\phi=0.
\ee

The generalised Garfinkle-Vachaspati transform takes the following form in the ten-dimensional string frame
 \bea
 g_{\mu\nu}&\to& g_{\mu\nu}+\Psi e^{-\phi}(k_\mu l_\nu+k_\nu l_\mu ),\label{GGV_main_text_1}\\
C&\to& C-\Psi e^{-2\phi} (k_\mu l_\nu-k_\nu l_\mu ).\label{GGV_main_text_2}
\eea
It is a valid solution generating technique provided 
\be
k^\mu F_{\mu\nu\rho}=-d(e^{-\phi}k)_{\nu\rho}, \label{transversality_main_text}
\ee
is satisfied by the background solution. We refer to this condition as the transversality condition. The scalar $\Psi$ should satisfy the following wave equation on the background spacetime,
\be
\Box\Psi-2(\partial_\mu\phi)g^{\mu\nu}(\partial_\nu\Psi)=0.
\ee
This equation can equivalently be written as
\be
\del_{\mu} (e^{-2\phi}g^{\mu\nu} \del_\nu\Psi)=0. \label{wave_eq_main_text}
\ee
In addition we also require that the scalar $\Psi$ is compatible with the Killing symmetries, i.e., 
\be
k^\mu\partial_\mu\Psi=0,\qquad \qquad
l^\mu\partial_\mu\Psi=0.
\ee

To establish the above statements we present a detailed calculation in appendix \ref{app:EOM_analysis}. There are two main steps involved in this computation. First, we do a conformal transformation such that the spacelike Killing vector $l^\mu$ becomes covariantly constant. Once this is achieved, we adapt technology from the previous paper \cite{GGV} to find the transformations of the left and right hand sides of the IIB equations of motion. We show that provided the transversality condition \eqref{transversality_main_text} and the wave equation \eqref{wave_eq_main_text} are satisfied, all ten-dimensional equations transform covariantly. Hence, we show the  generalised Garfinkle-Vachaspati transform 
\eqref{GGV_main_text_1}--\eqref{GGV_main_text_2}
is a valid solution generating technique with dilaton. We have also checked these computations independently using Cadabra \cite{Cadabra1,Cadabra2}.

For the embedding \eqref{metric_string_10d_main_text}, the ten-dimensional Einstein frame metric takes the form,
\be
ds_\rom{(E)}^2=e^{-\phi/2}ds_{6}^2+e^{\phi/2} ds_4^2\label{metric_einstein_10d_main_text}.
\ee
One can write the corresponding GGV in Einstein frame. Since for applications to the D1-D5 system in section \ref{sec:D1-D5} we only work with string frame metric, we relegate these details to appendix \ref{app:summary}.

\subsection{Transform for the NS-NS sector}
\label{sec:GGV_NS}

In our conventions the ten-dimensional NS-NS sector string frame action is,
\be
S_\rom{NS}=\frac{1}{16 \pi G_{10}}\int \sqrt{-g}e^{-2\Phi}\left[R+4(d\Phi)^2-\frac{1}{12}H_{\mu \nu \rho}H^{\mu \nu \rho}\right],
\ee
where $H=dB$.  The embedding of interest of six-dimensional theory \eqref{action_6d_main_text}
 in the ten-dimensional NS-NS sector string frame is,
\be
ds^2_{\rom{(S)}} = e^{-\phi} ds^2_6 + ds_4^2, \label{embedding_NS_string}
\ee
with  ten-dimensional dilaton
\be
\Phi = -\phi.
\ee 
The six-dimensional 2-form field is now the 2-form B-field with zero components in the four torus directions.

In Einstein frame this embedding reads
\be
ds^2_{\rom{(E)}} = e^{-\Phi/2} ds^2_{\rom{(S)}}  =e^{\phi/2} ds^2_{\rom{(S)}}  = e^{-\phi/2} ds^2_6 +e^{\phi/2} ds_4^2\label{embedding_NS_einstein_main_text}.
\ee
Note that this metric is same as \eqref{metric_einstein_10d_main_text}. In fact, the two embeddings are related by S-duality. S-duality relates the RR sector of IIB supergravity to the NS-NS sector. The S-duality transformation in Einstein frame is
\begin{align}
g^\rom{(E)}_{\mu\nu}&\to g^\rom{(E)}_{\mu\nu},& \Phi &\to -\Phi,& C_{\mu\nu}&\to B_{\mu\nu}.
\end{align}
The equivalence of \eqref{embedding_NS_einstein_main_text} and \eqref{metric_einstein_10d_main_text} is the reflection of the fact that the Einstein frame metric does not change under S-duality.

We can adapt the GGV from the RR sector to the NS-NS sector, for details see appendix \ref{app:summary}. In string frame the generalised Garfinkle-Vachaspati transform takes the form, 
\bea
g_{\mu\nu}&\to& g_{\mu\nu}+\Psi (k_\mu l_\nu + k_\nu l_\mu ),\\
B_{\mu\nu} &\to& B_{\mu\nu}-\Psi (k_\mu l_\nu - k_\nu l_\mu ).
\eea
The transversality condition reads,
\be
k^\mu  H_{\mu\nu\rho}=-(dk)_{\nu\rho},
\ee
and the scalar wave equation for the field $\Psi$ reads,
\be
\del_\mu (e^{-2\phi}g^{\mu\nu}\del_\nu\Psi)=0.
\ee

\section{Application to the D1-D5 system}
\label{sec:D1-D5}
In this section we explore applications of the GGV transform to a class of D1-D5 geometries. Consider type IIB string theory compactified on S$^1\times$ T$^4$. We denote by $y$ the coordinate of S$^1$ and by $z_i$ ($i=1,\ldots,4$) those of T$^4$. The radius of S$^1$ is $R_y$. The volume of T$^4$ is $(2\pi)^4\,V$.
\subsection{Deformation of a class of D1-D5-P backgrounds}
We consider traveling wave deformation along the torus directions  on the class of D1-D5-P backgrounds constructed in \cite{gms1, gms2}. A standard form \cite{gms2} for the string frame metric, RR 2-form and dilaton for this background is, 
\bea
\label{stringmetric}
ds^2 &=& -\frac{1}{h}\,(dt^2-dy^2) +\frac{Q_p}{h\,f}\,(dt-dy)^2 + h\,f\,\Bigl(\frac{dr^2}{r^2 + (\gamma_1+\gamma_2)^2\,\eta}+d\theta^2\Bigr)\nonumber\\
&&+ h\,\Bigl(r^2 + \gamma_1\,(\gamma_1+\gamma_2)\,\eta - \frac{Q_1 Q_5 \, (\gamma_1^2-\gamma_2^2)\,\eta\,\cos^2\theta}{h^2\,f^2}\Bigr)\,\cos^2\theta\,d\psi^2\nonumber\\
&&+ h\,\Bigl(r^2 + \gamma_2\,(\gamma_1+\gamma_2)\,\eta + \frac{Q_1 Q_5\, (\gamma_1^2-\gamma_2^2)\,\eta\,\sin^2\theta}{h^2\,f^2}\Bigr)\,\sin^2\theta\,d\phi^2\nonumber\\
&&+\frac{Q_p \, (\gamma_1+\gamma_2)^2\,\eta^2}{h\,f}\,(\cos^2\theta \,d\psi + \sin^2\theta\,d\phi)^2\nonumber\\
&&-\frac{2\,\sqrt{Q_1 Q_5}}{h\,f}\,(\gamma_1\,\cos^2\theta\,d\psi + \gamma_2\,\sin^2\theta\,d\phi)\,(dt-dy)\nonumber\\
&&-\frac{2\,\sqrt{Q_1 Q_5}\,(\gamma_1+\gamma_2)\,\eta}{h\,f}\,(\cos^2\theta\,d\psi + \sin^2\theta\,d\phi)\,dy + \sqrt{\frac{H_1}{H_5}}\,(dz^idz^i)\,,
\eea
\bea
C^{(2)} &=& - \frac{\sqrt{Q_1 Q_5}\, \cos^2\theta}{H_1\,f}\,(\gamma_2\,dt+\gamma_1\,dy)\wedge d\psi - \frac{\sqrt{Q_1 Q_5}\,\sin^2\theta}{H_1\,f}\,(\gamma_1\,dt+\gamma_2\,dy)\wedge d\phi\nonumber\\
&& + \frac{(\gamma_1+\gamma_2)\,\eta\,Q_p}{\sqrt{Q_1 Q_5}H_1 \,f}\,(Q_1\,dt+Q_5\,dy)\wedge (\cos^2\theta \,d\psi + \sin^2\theta\,d\phi)\nonumber\\
&&-\frac{Q_1}{H_1\,f}\,dt\wedge dy- \frac{Q_5\,\cos^2\theta}{H_1\,f}\,(r^2 +\, \gamma_2 \,(\gamma_1+\gamma_2)\,\eta + Q_1)\,d\psi\wedge d\phi\,,
\eea
\be
e^{2\Phi} = \frac{H_1}{H_5}\,,
\ee
where
\bea
\g_1 = - a m\,,\quad \g_2 = a \left(m + \frac{1}{k} \right)
\eea
and
\bea
&&a=\frac{\sqrt{Q_1\,Q_5}}{R_y}\,,\quad Q_p = -\,\gamma_1\,\gamma_2\,\quad \eta = \frac{Q_1\, Q_5}{Q_1\,Q_5 + Q_1\, Q_p + Q_5 \,Q_p}\,,\nonumber\\
&&f = r^2 + a^2\,(\gamma_1+\gamma_2)\,\eta\,(\gamma_1\,\sin^2\theta + \gamma_2\,\cos^2\theta)\,,\nonumber\\
&&H_1=1+\frac{Q_1}{f}\,,\quad H_5=1+\frac{Q_5}{f}\,,\quad h =\sqrt{H_1\,H_5}\,.
\eea
This configuration carries D1, D5, and P charges. The integer number of D1, D5, and P branes $n_1$, $n_5$, $n_p$, respectively are related to the parameters appearing in the metric as follows
\begin{align}
\label{Q1Q5}
Q_1&= \frac{g\,\alpha'^3}{V}\,n_1, & Q_5 &=g\,\alpha'\,n_5, & Q_p &=  \frac{g^2\,\alpha'^4}{V\,R_y^2}\,n_p.
 \end{align}

This above metric can be written in a generalised GMR form as a 2D fiber over a 4D almost hyper-K\"ahler base space. Note that we call it a generalised GMR form, as the nomenclature GMR form typically refers to supersymmetric solutions of minimal 6D supergravity \cite{GMR}. For non-minimal 6D supergravity supersymmetric solutions have been recently studied in \cite{Lam:2018jln, Cano:2018wnq}. As a 2D fibre over a 4D base space the string frame metric takes the form
\be
ds^2= -h^{-1}(dv + \beta)\Bigl(du + \omega + \frac{\cF}{2}(dv + \beta)\Bigr) + h h_{mn} dx^m dx^n + \sqrt{\frac{H_1}{H_5}} (dz^idz^i),
\label{general6d}
\ee
where $u = t+y$ and $ v=t-y$, and\footnote{In the previous paper \cite{GGV} $\cF$ was mistakenly written to be $\cF = -\frac{Q_p}{f}$; correct expression is $\cF = -\frac{2Q_p}{f}$.}
\bea
\cF &=& -\frac{2Q_p}{f},\\
 \beta&=& \frac{\sqrt{Q_1 Q_5}}{f}\,(\g_1 + \g_2)\,\eta\,(\cos^2\theta\,d\psi
+ \sin^2\theta\,d\phi), \\
 \omega&=& \frac{\sqrt{Q_1 Q_5}}{f}\,\Bigl[\Bigl(2\g_1 -
(\g_1 + \g_2)\,\eta\,\Bigl(1-2 \frac{Q_p}{f}\Bigr)\Bigr)\,\cos^2\theta\,d\psi \nn \\
&&  \quad + \Bigl(2\g_2 - (\g_1 + \g_2)\,\eta\,\Bigl(1-2
\frac{Q_p}{f}\Bigr)\Bigr)\,\sin^2\theta\,d\phi\Bigr],
\label{3chargegmr}
\eea
and the base metric $h_{mn}$ given as,
\bea
ds^2_\rom{base} &=& h_{mn} dx^m dx^n= f\left(\frac{dr^2}{r^2+(\g_1
+ \g_2)^2\, \eta} + d\theta^2\right)\nonumber\\
&&+ \frac{1}{f}\Bigl[[r^4 +
r^2\,(\g_1+\g_2)\,\eta\,(2\g_1 - (\g_1-\g_2) \cos^2\theta) +
(\g_1+\g_2)^2\,\g_1^2\,\eta^2\,\sin^2\theta]\cos^2\theta\,d\psi^2
\nn \\
&&+[r^4 +
r^2\,(\g_1+\g_2)\,\eta\,(2\g_2 + (\g_1-\g_2) \sin^2\theta) +
(\g_1+\g_2)^2 \,\g_2^2 \,\eta^2 \,\cos^2\theta]\sin^2\theta
\,d\phi^2\nn \\
&&-2\g_1 \g_2 \,(\g_1+\g_2)^2\, \eta^2\,\sin^2\theta \cos^2\theta \,d\psi d\phi\Bigr].
\label{3chargebase}
\eea

The above configuration has 
\be
k = \frac{\partial}{\partial u},
\ee
as the appropriate null Killing vector and 
\be
l^{(i)} = \frac{\partial}{\partial z^i},
\ee
as the appropriate spacelike Killing vector for the application of the generalised Garfinkle-Vachaspati transform.  The background configuration also satisfies the transversality condition, cf.~\eqref{transversality_main_text}, 
\be
k^\mu F_{\mu \nu \rho} = - (d (e^{-\Phi} k))_{\nu \rho}.
\ee

A general solution to the scalar equation, cf.~\eqref{wave_eq_main_text}, 
\be
\partial_\mu \left[ e^{-2\Phi} \sqrt{-g} g^{\mu \nu} \partial_\nu \Psi \right] = 0,
\ee
can be obtained using the ansatz
\be
\Psi = \sum_{n=-\infty}^{\infty}f_n(r) \exp \left[-i n\frac{v}{R_y} \right].
\ee
Upon substituting this ansatz we get ordinary differential equations for the functions $f_n(r)$, which can be readily solved. 
We find
\be
\Psi_i (r,v) =
 \sum_{n=-\infty}^{\infty}c_n^i \, \left(\frac{r^2}{r^2 \left( 1 +  a^2 \frac{(Q_1 + Q_5)}{ Q_1 Q_5}m\left(m+\frac{1}{k}\right)\right) + \frac{a^2}{k^2}}\right)^{\frac{|n|k}{2}} e^{-in\frac{v}{R_y}}. \label{scalar_m}
\ee
where the index $i$ refers to the four-torus coordinates $z^i$.

The generalised Garfinkle-Vachaspati transform
\bea
g_{\mu \nu} &\to& g_{\mu \nu} + \sum_{i=1}^{4}\Psi_i e^{-\Phi} (k_\mu l_\nu^{(i)} + k_\nu l_\mu^{(i)}), \\
C_{\mu \nu} &\to& C_{\mu \nu} - \sum_{i=1}^{4}\Psi_i e^{-2\Phi} (k_\mu l_\nu^{(i)} - k_\nu l_\mu^{(i)}), 
\eea
simply corresponds to  replacing
\be
du \to du + \Psi_i dz^i,
\ee
in the metric. For the form field, the recipe is the same, but there are some details. The two form field written above can also be written as 
\bea
C^{(2)} &=& -\frac{1}{2} \left(H_1^{-1}\right) du \wedge dv \nn \\
 && + \,(\gamma_1+\gamma_2)\frac{(\eta \, Q_p \, (Q_1+ Q_5) - Q_1 \, Q_5)}{2\sqrt{Q_1 \, Q_5} \, H_1 \,f}\,  du \wedge (\cos^2\theta \,d\psi + \sin^2\theta\,d\phi)\nn \\
&& +\,(\gamma_1+\gamma_2)\, \frac{\eta \, Q_p \, (Q_1- Q_5)}{2\sqrt{Q_1 \,Q_5} \, H_1 \,f} dv \wedge (\cos^2\theta \,d\psi + \sin^2\theta\,d\phi)\nonumber\\
&& + \,(\gamma_1-\gamma_2)\,\frac{\sqrt{Q_1\, Q_5}}{2\,H_1 \,f}\, dv \wedge (\cos^2\theta \,d\psi - \sin^2\theta\,d\phi)\nonumber\\
&& - \frac{Q_5\,\cos^2\theta}{H_1\,f}\,(r^2 + \gamma_2 \,(\gamma_1+\gamma_2)\,\eta + Q_1)\,d\psi\wedge d\phi\,,
\eea
where we have removed a constant term proportional to  $du \wedge dv$  by a gauge transformation. In this form the deformation of  $C^{(2)} $ also simply corresponds to  replacing
\be
du \to du + \Psi_i dz^i.
\ee
When $Q_1 = Q_5$ the deformation reduces to to the one considered in \cite{GGV}.

\subsection{Global properties and smoothness} 
Although it is not manifest in the above coordinates, the deformed solution has flat asymptotics~\cite{Lunin:2012gp, GGV}. Much of the following discussion in this section parallels the corresponding discussions in those references, so we shall be brief.  At infinity the metric of the deformed solution looks like,
\be
ds^2=-[du+f_i(v)dz^i]dv+dr^2+r^2 d\Omega_3^2+dz^i dz^i,
\label{asymptotic_deformed}
\ee
where
\be
f_i(v) = \lim_{r\to \infty} \Psi_i(r,v) = \sum_{n \neq 0} c_n^i \left(1 +  \frac{a^2(Q_1+Q_5)}{ Q_1Q_5}m\left(m+\frac{1}{k}\right)\right)^{-\frac{|n|k}{2}} e^{-in\frac{v}{R_y}}. \label{f_i}
\ee

As in previous works \cite{Lunin:2012gp, GGV}, for simplicity, from now onwards we assume $c_0^i =0$. 
The coordinate transformation that puts the deformed spacetime in an asymptotically flat form and simplifies the extraction of charges is
\bea
z'{}_i &=& z{}_i - \frac{1}{2} \int_0^v \Psi_i( r,\tilde v) d \tilde v,  \label{diff1} \\ 
u' &=&\lambda \left[ u + \frac{1}{4}\int_0^{v} \Psi_i (r, \tilde v)\Psi_i( r, \tilde v) d \tilde v\right], \\
v' &=& \frac{v}{\lambda}. \label{diff3}
\eea
In the $r \to \infty$ limit, this transformation simplifies to,
\bea
z'{}_i &=& z{}_i - \frac{1}{2} \int_0^v f_i( \tilde v) d \tilde v, \label{asym_diff1} \\ 
u' &=&\lambda \left[ u + \frac{1}{4}\int_0^{v} f_i ( \tilde v)f_i( \tilde v) d \tilde v\right],  \label{asym_diff2}  \\
v' &=& \frac{v}{\lambda},\label{asym_diff3}
\eea
where
\be
\lambda^{-2} =  1 - \frac{1}{8 \pi R_y} \int_0^{2 \pi R_y} f_i( \tilde v) f_i( \tilde v) d \tilde v. \label{lambda2}
\ee
The value of $\lambda$ is fixed by the requirement that the new time coordinate $t' = \frac{1}{2}(u' + v')$
 is a single valued function under $y \sim y + 2 \pi R_y$ at infinity. In new coordinates, the asymptotic metric \eqref{asymptotic_deformed} is manifestly flat,
\be
ds^2 = - (dt')^2 + (dy')^2 + dr^2 + r^2 d \Omega^2_3 + dz'{}^i dz'{}^i.
\ee
The $z'{}^i$ coordinates have the same periodicity as the $z{}^i$ coordinates. The periodicity of the $y' = \frac{1}{2}(u' + v')$ coordinate is $y' \sim y' + 2 \pi R$, with  $ R =  \lambda^{-1} R_y.$ In the rest of the section we  exclusively work with $R$ as opposed to $R_y$.

Following \cite{Lunin:2012gp}, we introduce
\be
h_i(v') := f_i(v) = f_i(\lambda v').
\ee
In terms of the parameter $R$ we have,
\be
\lambda^{-2} = 1 - \frac{1}{8 \pi R} \int_0^{2 \pi R} h_i( \tilde v') h_i( \tilde v') d \tilde v'. \label{lambda}
\ee

Now we would like to extract the ADM quantities. We find that it is most convenient to do this computation in six-dimensions as the ten-dimensional string frame metric is directly related to the six-dimensional Einstein frame metric. At large values of $r$ we find that the relevant terms of the ten-dimensional string frame metric admits an expansion of the form,
 \begin{align}
&g_{t't'} = -1 + \frac{1}{r^2} \left( \frac{Q_1+Q_5}{2}+ \lambda^2 Q_p + \frac{1}{4} \lambda^2 Q_1 h_i h_i\right) + \ldots\\
&g_{t'y'}= - \frac{\lambda^2}{r^2} \left( Q_p + \frac{1}{4} Q_1 h_i h_i  \right)+ \ldots\\
&g_{y'y'}= 1+ \frac{1}{r^2} \left( -\frac{Q_1+Q_5}{2}+ \lambda^2 Q_p + \frac{1}{4} \lambda^2 Q_1 h_i h_i\right)+ \ldots\\
&g_{t'\phi}=- \frac{\lambda \sqrt{Q_1Q_5}}{r^2} s_\theta^2 \left( \gamma_2 - \frac{\gamma_1 + \gamma_2}{2} \eta \left( 1- \frac{1}{4}h_i h_i - \frac{1}{\lambda^2}\right)\right)+ \ldots\\
&g_{t'\psi}=-\frac{\lambda \sqrt{Q_1Q_5}}{r^2} c_\theta^2 \left( \gamma_1 - \frac{\gamma_1 + \gamma_2}{2} \eta \left( 1- \frac{1}{4}h_i h_i - \frac{1}{\lambda^2}\right)\right)+ \ldots \;\;.
\end{align}
From these components we can extract (six-dimensional) ADM quantities. The ADM momenta in the $y'$-direction is
\bea
P_{y'} &=& - \frac{\pi}{4 G} \int_0^{2\pi R} dy \ r^2 \ \delta g_{t'y'} =  \frac{\pi \lambda^2}{4G} \left( 2 \pi R \ Q_p + \frac{1}{4} Q_1 \int_{0}^{2\pi R} dy' h_i h_i   \right) \\
 &=& \frac{n_1 n_5}{R} \left[ m \left( m + \frac{1}{k} \right)+ \frac{Q_1}{4a^2} \frac{1}{2\pi R} \int_0^{2\pi R} dy' h_i h_i \right], \label{charge_py}
\eea
where we have used the 6-dimensional Newton's constant to be  $G=\frac{\pi^2\alpha'^4g^2}{2V}$ together with \eqref{Q1Q5}. The ADM mass is \cite{GGV}
\bea
M &=&  \frac{\pi}{8 G} \int_0^{2\pi R} dy \ r^2 \ (3 \delta g_{t't'}  - \delta g_{y'y'}) \\
&=&  \frac{\pi}{4G} (Q_1+Q_5) (2 \pi R) + P_{y'}.
\eea
We note that the BPS bound is saturated; addition of momentum shifts the mass by $P_{y'}$. 
To extract angular momenta, we use
\bea
J_\phi &=& - \frac{\pi}{8 G} \int_0^{2 \pi R} dy' \ r^2 \frac{\delta g_{t'\phi}}{\sin^2 \theta}  = \frac{n_1 n_5}{2} \left( m + \frac{1}{k}\right), \label{charge_Jphi} \\
J_\psi &=& - \frac{\pi}{8 G} \int_0^{2 \pi R} dy' \ r^2 \frac{\delta g_{t'\psi}}{\cos^2 \theta} =  - \frac{n_1 n_5}{2}  m.\label{charge_Jpsi}
\eea

To analyse the smoothness of the spacetime, we start by looking at the determinant of the deformed metric. The determinant  of the deformed metric remains the same as the undeformed metric. Furthermore, since the scalar \eqref{scalar_m} is finite everywhere, potential singularities can only occur at places where the background solution becomes singular. In the background solution, there are no such points~\cite{gms1,gms2}. Hence the solution remains smooth even after the deformation.

\subsection{Decoupling limit and AdS/CFT interpretation}
The {\emph{undeformed}} geometry develops a large AdS region when,
\be
\epsilon \equiv \frac{\sqrt{Q_1 Q_5}}{R_y^2} = \frac{a^2}{\sqrt{Q_1 Q_5}} \ll 1.
\ee
To obtain the decoupled metric we introduce,
\begin{align}
\bar u &= \frac{u}{R_y}, &
\bar v &= \frac{v}{R_y}, &
\bar r &= \frac{r}{a}, \label{rescaling}
\end{align}
and take the limit $R_y \to \infty$ keeping $Q_1$ and $Q_5$ fixed. Due to rescaling of coordinates \eqref{rescaling} we get a metric that describes the inner AdS$_3~\times$~S$^3~\times$~T$^4$ region of the geometry,
\bea 
ds^2&=&\sqrt{Q_1 Q_5} \left[-{\bar r}^2d{\bar u}d{\bar v}-
\frac{1}{4}(d{\bar u}+d{\bar v})^2+
\frac{d{\bar r}^2}{{\bar r}^2+k^{-2}}\right] \nn \\
&&+ \ \sqrt{Q_1 Q_5} \left[d\theta^2+
c_\theta^2\left(d\psi-\frac{1}{2k}(d{\bar u}-
d{\bar v})+md{\bar v}\right)^2+
s_\theta^2\left(d\phi-\frac{1}{2k}(d{\bar u}+
d{\bar v})-md{\bar v}\right)^2\right] \nn \\
& & + \sqrt{\frac{Q_1}{Q_5}}\ dz^i dz^i \,. \label{decoupled} 
\eea

To obtain the decoupled metric with the deformation turned on we proceed as follows~\cite{Lunin:2012gp, GGV}. In order to maintain ADM momentum \eqref{charge_py} finite as $R_y $ becomes large, we must scale the scalars appropriately with $R_y$.  In the present set-up, scalars should scale  as 
\be
\Psi_i =: \frac{a}{\sqrt{Q_1}} \,\bar \Psi_i = \frac{\sqrt{Q_5}}{R_y} \,\bar \Psi_i . 
\ee 
With this rescaling, terms in the metric of the form
$
du + \Psi_i dz^i
$ 
behave as 
\be
du + \Psi_i dz^i = R_y \, d \bar u + \frac{\sqrt{Q_5}}{R_y} \, \bar \Psi_i \, dz^i.
\ee
In the limit  $R_y \to \infty$ such terms simply become
$
R_y \, d \bar u,
$
i.e., scalars $\Psi_i$ all scale out. Once again we get the decoupled metric \eqref{decoupled}.

However, all is not that simple. There is a subtlety. We recall from the analysis of the previous subsection that the deformed metric is not manifestly asymptotically flat in coordinates $z_i, t, y$; it is manifestly asymptotically flat in  $z'_i, t', y'$. The decoupled metric in the $z'_i, t', y'$ coordinates is naturally glued to the asymptotically flat region. Therefore, we should write the decoupled metric in these coordinates. This change of coordinates reintroduces scalars.   In the $R_y \to \infty$ limit transformations \eqref{diff1}--\eqref{diff3} simplify to 
\begin{align}
z'_i &= z_i - \frac{1}{2} \sqrt{Q_5}  \int_0^{\bar v} \bar \Psi_i \, d \bar{ \tilde{v}}, & {\bar u}' &= \bar u ,&
{\bar v}' &= \bar v.
\end{align} 
The decoupled metric takes the form, 
\bea
ds^2&=&\sqrt{Q_1 Q_5}\left[-{\bar r}^2d{\bar u}d{\bar v}-
\frac{1}{4}(d{\bar u}+d{\bar v})^2+
\frac{d{\bar r}^2}{{\bar r}^2+k^{-2}}\right] \nn \\
&&+ \ \sqrt{Q_1 Q_5}\left[d\theta^2+
c_\theta^2\left(d\psi-\frac{1}{2k}(d{\bar u}-
d{\bar v})+md{\bar v}\right)^2+
s_\theta^2\left(d\phi-\frac{1}{2k}(d{\bar u}+
d{\bar v})-md{\bar v}\right)^2\right] \nn \\
& & + \ \sqrt{\frac{Q_1}{Q_5}} \ \left(dz'_i + \frac{1}{2} \sqrt{Q_5} \ \bar \Psi_i d \bar v \right)^2 \,. \label{decoupled2} 
\eea

We can now read off the charges, say, by comparing the above metric to a standard form of asymptotic form of the AdS$_3~\times$~S$^3~\times$~T$^4$. We find,
\bea
 P_{y'} &=& \frac{n_1 n_5}{R} \left[ m \left( m + \frac{1}{k} \right)+ \frac{1}{8\pi} \int_0^{2\pi} d\bar y \bar f_i \bar f_i \right],   \label{chargesAdS1}\\ 
J_\phi &=& \frac{n_1 n_5}{2} \left( m + \frac{1}{k}\right),  \label{chargesAdS2} \\
J_\psi &=& - \frac{n_1 n_5}{2}  m. \label{chargesAdS3}
\eea
These charges agree with expressions \eqref{charge_py}, \eqref{charge_Jphi}, \eqref{charge_Jpsi} in the $R_y \to \infty$ limit.

\subsection{Deformed states in the D1-D5 CFT}
Let $|\psi \rangle$ be the normalised state in the D1-D5 CFT that describes the dual to the undeformed gravity configuration with ADM momentum
\be
 \frac{n_1 n_5}{R} \left[ m \left( m + \frac{1}{k} \right) \right].
 \ee 
 Then the expression for the momentum $P_{y'}$, cf.~\eqref{chargesAdS1}, can be compared with the momentum of the normalised deformed CFT state~\cite{Lunin:2012gp, GGV}, 
\be
| \Psi \rangle = \exp \left[ - \frac{n_1 n_5}{4} \sum_{n>0} n  (\mu^i_n)^* \mu^i_n\right] \exp\left[ \sum_{n>0} \mu_n^i J_{-n}^i\right] |\psi \rangle,
\ee
where  $J_{-n}^i$ are the modes of the four U(1) currents of the D1-D5 CFT, and the parameters $\mu_n^i$  are determined below.
The momentum of the deformed state turns out to be,
 \bea
 R P_{y'}  &:=&  \langle \Psi | L_0 - \bar L_0 | \Psi \rangle =  \langle \psi | L_0 - \bar L_0 | \psi \rangle +  \sum_{ n> 0} \frac{n^2 n_1 n_5}{2}( \mu^i_n)^* \mu^i_n \\
 &=&n_1 n_5 \left[ m \left( m + \frac{1}{k} \right) \right] +  \sum_{ n> 0} \frac{n^2 n_1 n_5}{2}( \mu^i_n)^* \mu^i_n
\eea
Upon doing the Fourier expansion of \eqref{chargesAdS1} in the decoupling limit, we get (assuming $\Psi_i$'s are real scalars),
\be
R P_{y'} = n_1 n_5 \left[ m \left( m + \frac{1}{k} \right) \right]  +   \sum_{ n> 0} \frac{n_1 n_5}{2} \frac{Q_1}{a^2}\left(  (c^i_n)^* c^i_n \right).
\ee
From the matching between the gravity and the CFT answers we arrive at the relation between the quantities $c_n^i$ and $\mu_n^i$,
\be
\mu_n^i  = \frac{1}{n} \frac{\sqrt{Q_1}}{a}c_n^i. 
\ee
With this identification we have singled out a state in the  D1-D5 CFT that has the same charges as the deformed gravity solutions. 

\section{Application to the F1-P system}
\label{sec:F1-P}

In this section, we briefly discuss application of the GGV transform to the  F1-P system. Since the most general vibrating  fundamental string solution with momentum modes added on top is already well known~\cite{Dabholkar:1995nc, Callan:1995hn}, we do not expect that the GGV technique would allow us to discover something novel. Nonetheless, the F1-P system is well suited for an application of the GGV transform in the NS sector.\footnote{Application to the NS1-NS5 bound states is also a possibility, but this set-up is related to the D1-D5 set-up by S-duality.}

We start with the chiral null model for the NS sector of type II supergravity in Einstein frame \cite{Horowitz:1994ei, Horowitz:1994rf}. We take $\partial_u$ as the null Killing vector. Metric and the supporting matter fields take the form,
\bea \label{chiral_null_1}
&& ds^2 = H^{3/4} (-dudv + K dv^2 + 2A_i dx^i dv) + H^{-1/4}(dx^i dx^i + dz^j dz^j), \label{chiral_null_metric} \\
\label{chiral_null_2}
&& B_{uv} = \frac{H}{2} \ ,  \qquad B_{vi} = -H A_i \ , \qquad  e^{2\phi} = H , \label{chiral_null_matter}
\eea
where $x^i$ with $i=1,\ldots, 4$ are non-compact cartesian coordinates on  $\mathbb{R}^4$ and $z^j$ with $j=1,\ldots,4$ are the T$^4$ coordinates. The  supergravity equations of motion are satisfied provided $H^{-1}$ and $K$ are  harmonic functions on the transverse space $\mathbb{R}^4~\times~$T$^4$ parameterised by  $(x^i, z^j)$. Upon smearing these functions on T$^4$, they   become harmonic functions on $\mathbb{R}^4$.  The coefficients and the sources for these harmonic functions can depend on $v$. The functions $A_i$ can be thought of as a gauge field. The chiral null model equations of motion require that it satisfies source-free Maxwell equation, see e.g., appendix C of \cite{Lunin:2001jy}.  In general the gauge field can also have components in the T$^4$ directions; in the above metric we have written only for the $\mathbb{R}^4$ directions. Let $u=t + y, v = t-y$ with the coordinate $y$ be periodic with length $L_y = 2 \pi R_y$. Level matched  F1-P configurations with smeared harmonic functions  over the four-torus and the $y$ direction are described by~\cite{Lunin:2001jy},
\begin{align}
H^{-1} &= 1+ \frac{Q}{L_y} \int_0^{L_y} \frac{dv}{|x-F(v)|^2}, & A_i &= -\frac{Q}{L_y} \int_0^{L_y} \frac{dv \dot F_i(v)}{|x-F(v)|^2}, & K &=  \frac{Q}{L_y} \int_0^{L_y} \frac{dv \dot F_i(v)\dot F_i(v)}{|x-F(v)|^2},& \label{chiral_null_3}
\end{align}
where $|x-F(v)|^2 = \sum_{i=1}^{4} (x_i - F_i(v))^2$.

We wish to apply the generalised Garfinkle-Vachaspati transform to such a smeared chiral null model solution, with $k^\mu=(\partial_u)^\mu$ as the null Killing vector and $l^{\mu}_{(i)}= (\partial_{z^i})^{\mu}$ as the spacelike Killing vector. Under the GGV transform, Einstein frame metric  and the $2$-form field transform as (see appendix \ref{app:summary}),
 \bea
 g_{\mu\nu} &\to& g_{\mu\nu} + \Psi_{(j)} e^{\phi/2}(k_\mu l^{(j)}_\nu + k_\nu l^{(j)}_\mu) \ , \\
 B_{\mu\nu} &\to& B_{\mu\nu} - \Psi_{(j)} e^{\phi}(k_\mu l^{(j)}_\nu - k_\nu l^{(j)}_\mu)\ ,
  \eea
 where functions $\Psi_{(j)}$ satisfy $\Box \Psi_{(j)} =0$ for $j=1,\ldots,4$.  Applying this transformation to configuration \eqref{chiral_null_metric}--\eqref{chiral_null_matter} we get the transformed metric and $2$-form field as
 \bea
 ds^2 &=& H^{3/4} (-du dv + K dv^2 + 2A_i dx^i dv - \Psi_{(j)}dz^j dv) + H^{-1/4}(dx^i dx^i + dz^j dz^j) \label{interpret:GGV1} \\
  B_{uv} &=& \frac{H}{2} \ , \qquad  B_{vi}~=~-H A_i \ , \label{interpret:GGV2} \\
  B_{vz_j}&=& H \, \Psi_{(j)}  \ , \qquad e^{2\phi}~=~H \, .
\label{interpret:GGV3}
\eea
The harmonic functions $\Psi_{(j)}$ in equations \eqref{interpret:GGV1}--\eqref{interpret:GGV3} can be interpreted as additional components of the gauge field $A_i$ with components in the torus directions, see, e.g., \cite{Lunin:2002iz, Kanitscheider:2007wq}. The final solution is also a chiral null model solution.  In general, it does not satisfy the level matching condition~\cite{Callan:1995hn}. One can always perform an ordinary GV transform to add appropriate momentum to get  a  solution that satisfies the level matching condition.

Alternatively, taking 
\be
\Psi_{(j)}(v,x_i)= -2H^{-1} p_{j}(v),
\ee  
we can arrive at a slightly different interpretation as follows.
The transformed metric takes the form,
\be
ds^2 = H^{3/4} (-dudv + K dv^2 + 2A_i dx^i dv ) + H^{-1/4}(dx^i dx^i )+ H^{-1/4}(dz^j dz^j  + 2p_j dz^j dv).
\ee
Introducing new coordinates $z'{}^j= z^j + \int_{0}^{v} p^j (v')dv'$, we can write the above metric as, 
\be
ds^2 = H^{3/4} (-dudv + K dv^2 + 2A_i dx^i dv -H^{-1} p^2(v) dv^2) + H^{-1/4}(dx^i dx^i + dz'{}^j dz'{}^j) .
\ee
The two-form B-field and the dilaton remain unchanged under this coordinate transformation,
\be
B_{uv} = \frac{H}{2} \ , \ B_{vi}~=~-H A_i \ , \   B_{vz'{}^j}= -2p_j (v) \ , \  e^{2\phi}~=~H.
\ee
The component $B_{vz'{}^j}= -2p_j (v)$ can be removed by a gauge transformation $B'_{ab} = B_{ab} + \partial_a \Lambda_b  - \partial_b \Lambda_a$ with gauge function 
$\Lambda_{z'{}^j} = 2\int_{0}^{v} p_j (v')dv'$ (no other component changes  under this gauge transformation). We finally have 
\bea
ds^2 &=& H^{3/4} (-dudv + K dv^2 + 2A_i dx^i dv -H^{-1} p^2(v) dv^2) + H^{-1/4}(dx^i dx^i + dz'{}^j dz'{}^j), \label{interpret:GGV_GV} \\
 B_{uv} &=& \frac{H}{2} \ , \qquad  B_{vi}~=~-H A_i \  ,  \  \qquad e^{2\phi}~=~H.
\eea
This  metric can be readily interpreted as pp-wave  added to the above F1-P chiral null model with matter fields remaining unchanged.\footnote{This is similar to the interpretation given in section $7.1$ of reference \cite{Lunin:2012gp}, though for a different set-up where matter fields do change.} The final metric is simply the ordinary GV transform~\eqref{ordinary_GV1}-\eqref{ordinary_GV3} on the F1-P chiral null model \eqref{chiral_null_1}-\eqref{chiral_null_3} with harmonic function $\chi = H^{-1}$.  In the terminology of  \cite{Dabholkar:1995nc}, this choice of the harmonic function $\chi$ in the GV transform corresponds to adding ``momentum waves without oscillations.'' Due to the constant term in the harmonic function $H^{-1}$,  metric \eqref{interpret:GGV_GV}  is not manifestly asymptotically flat. It can be made asymptotically flat by shifting $u$ appropriately. In general, the final metric does not satisfy the level matching condition.

\section{Summary and future directions}
\label{sec:disc}
In this paper, we presented a generalised Garfinkle-Vachaspati transform for a six-dimensional theory with dilaton embedded in ten dimensions via the addition of four-torus. We considered two different embeddings, related by S-duality. In both these embeddings we presented the GGV technique  and worked out its applications. 

In the type IIB Ramond-Ramond embedding,  the technique allows us to add travelling-wave deformations  involving the torus directions on a class of D1-D5-P geometries. We explicitly worked out these deformations on certain supersymmetric D1-D5-P orbifolds. The deformed solutions are given in terms of solutions of a scalar field on the background geometry. When the background contains a large AdS region, the deformed states are identified in the D1-D5 CFT as an action of a U(1) current on the undeformed state. This analysis is a generalisation of the previous works \cite{Lunin:2012gp, GGV}, where similar deformations were considered for the dilaton free solutions. Using our techniques, deformations of further examples can be considered, including microstates of the D1-D5-KK system~\cite{Bena:2005ay}. We also briefly discussed the application of the GGV technique to the F1-P system.

Our work offers several opportunities for future research. The most natural direction to explore is  the supersymmetry properties of the deformed solutions.  In \cite{Lunin:2012gp} it was shown  that under the GGV transform, supersymmetric solutions of minimal six-dimensional supergravity are deformed into supersymmetric solutions of ten-dimensional IIB supergravity.  It is natural to conjecture that the GGV deformations of supersymmetric solutions of non-mininal  six-dimensional supergravity are also supersymmetric in ten dimensions. 

In a related line of investigation, reference \cite{Lam:2018jln} studied a class of BPS black string solutions with traveling waves. The horizon of these solutions turns out to be singular. It will be interesting to understand if our technique allows one to  add non-singular travelling wave deformations on black string solutions lifted to ten-dimensions. Such hair will also be of interest with regard to the 4D-5D connection in IIB compactification on  T$^4 \times $ S$^1$, cf.~\cite{Banerjee:2009uk, Jatkar:2009yd}. 
 
  It will be useful to  explore GGV technique in other duality frames, in particular, say for solutions of five-dimensional STU supergravity embedded in M-theory. The structure of embedding, see, e.g.,~\cite{Virmani:2012kw}, is very similar as in the present work, though because of the presence of different matter fields details are likely to be different.  If successfully implemented, the technique will allow us to add hair modes associated to various U(1) currents of the MSW CFT  on the MSW microstates \cite{Bena:2017geu}.

\subsection*{Acknowledgements}
We thank Dileep Jatkar and Ashoke Sen  for discussions and encouragement. AV thanks NISER Bhubaneshwar for kind hospitality towards the final stages of this work. The work of AV is supported in part by Max-Planck Partner Group ``Quantum Black Holes'' between CMI Chennai and AEI Potsdam. We also thank the anonymous referees for suggesting improvements on an earlier version of the manuscript.

\appendix

\section{Detailed analysis of equations of motion}
\label{app:EOM_analysis}
In this appendix we establish that the generalised Garfinkle-Vachaspati (GGV) transform with dilaton is a valid solution generating technique. In the main text we presented the technique in the string frame.  For the purpose of showing that it is a valid solution generating technique it is most convenient to work in the so-called conformal frame (introduced below). Once we establish the technique, it can be readily written in any frame we like. We provide a concise summary in string, Einstein, and what we call conformal frame in section \ref{app:summary}.

To establish the solution generating technique we follow the same methodology as in our previous work \cite{GGV}. Via a detailed brute force calculation we show that the left hand side and the right hand side of the IIB equations transform in the same way. 
 
 The set-up we are interested in is six-dimensional metric, antisymmetric 2-form field $C_{\mu \nu}$, and dilaton $\phi$ lifted to ten-dimensional IIB theory on T$^4$. The six-dimensional set-up  is minimal six-dimensional supergravity coupled to one self-dual tensor multiplet with action~\cite{Duff}
 \be
 S_6 = \frac{1}{16\pi G_6} \int d^6x \sqrt{-g} \left[ R - (d\phi)^2  - \frac{1}{12} e^{2 \phi} F_{\mu \nu \rho}F^{\mu \nu \rho}\right], \label{action_6d}
 \ee
 For ten-dimensional fields we follow Polchinski's conventions. The
ten-dimensional IIB string frame action with RR 2-form field is 
\be
S_\rom{RR}=\frac{1}{16 \pi G_{10}}\int d^{10}x\sqrt{-g}\left[e^{-2\Phi}[R+4(d\Phi)^2]-\frac{1}{12}F_{\mu \nu \rho}F^{\mu \nu \rho}\right].
\ee
 The embedding of interest  of six-dimensional fields in ten-dimensions is
\be
ds^2_{\rom{(S)}}=ds_{6}^2+e^{\phi} ds_4^2,\label{metric_string_10d}
\ee
where $ds^2_{\rom{(S)}}$ is the ten-dimensional string frame metric, $ds^2_4 = \sum_{i=1}^{4} dz^i dz^i$ is the flat torus metric, $\phi$ is the six-dimensional dilaton. Ten-dimensional dilaton is same as the six-dimensional dilaton 
\be
\Phi =\phi,
\ee 
and the ten-dimensional 2-form R-R field is also same as the six-dimensional 2-form field with zero components in the four torus directions. The ten-dimensional Einstein frame metric takes the form,
\be
ds_\rom{(E)}^2=e^{-\phi/2}ds_{6}^2+e^{\phi/2} ds_4^2.\label{metric_einstein_10d}
\ee
In a standard IIB conventions~\cite{Pol1,Pol2}, bosonic field equations in Einstein frame  take the form 
\begin{align}
&R_{\mu \nu}=\frac{1}{2}\nabla_\mu\Phi\nabla_\nu\Phi+\frac{1}{4}e^{\Phi}\left(F_{\mu \rho \sigma}{F_\nu}^{\rho \sigma}- \frac{1}{12}g_{\mu \nu}F_{\rho \sigma \kappa}F^{\rho \sigma \kappa}\right),\label{einstein_eq}\\
&\del_{\mu}\left(e^{\Phi}F^{\mu \rho \sigma}\right) = 0,& \label{eq_matter} \\
&\Box \Phi = \frac{1}{12} e^{\Phi} F_{\rho \sigma \kappa}F^{\rho \sigma \kappa}.&\label{dilaton_eq}
\end{align}

\subsection{Left hand side of Einstein equations}
 Generalised Garfinkle-Vachaspati transform technique uses a null Killing vector and a spacelike Killing vector. 
In order to directly use some of the details from the previous paper \cite{GGV}, we need to do a conformal transformation when dilaton is present. This is because in the previous analysis we worked with a covariantly constant spacelike vector $l^\mu$ provided by the torus directions.  In neither the Einstein frame nor the string frame torus Killing vectors are covariantly constant. 

To obtain the form of the metric where the four torus directions provide covariantly constant vectors we perform a conformal transformation by the factor $e^{-\phi/2}$ on the Einstein frame metric. We call the new metric the ``conformal frame'' metric. It takes the form,
\be
ds^2_\rom{(C)} = e^{-\phi/2}ds_\rom{(E)}^2=e^{-\phi}ds_{6}^2+ds_4^2=\tilde{g}_{\mu\nu}dx^\mu dx^\nu\label{conformal_metric}.
\ee

Let the six-dimensional metric $ds^2_6$, and hence the ten-dimensional string frame metric $ds^2_\rom{(S)}$ cf.~\eqref{metric_string_10d}, admits a null Killing vector $k^\mu$,
with the property that the dilaton is compatible with the Killing symmetry
\be
k^\mu\partial_\mu\phi=0.
\ee
It then immediately follows that the conformal frame metric $ds^2_\rom{(C)}$ also admits the same Killing vector. Since the covariant (lower) components of the Killing 1-form depend on the metric under consideration, to avoid any confusion, we use the notation $\tilde k_\mu$ for the conformal frame Killing 1-form. We also use the notation $\tilde k^\mu$, but we note that $\tilde k^\mu  \equiv k^\mu$. This is a convenient notation. It follows that
\be
0=(\mathcal{L}_{\tilde{k}}\tilde{g})_{\mu\nu}=\tilde{\del}_\mu \tilde{k}_\nu +\tilde{\del}_\nu \tilde{k}_\mu,
\ee
where $\tilde{\del}_\mu$ is the metric compatible covariant derivative with respect to the conformal frame metric.

In the conformal frame the torus directions provide covariantly constant unit normalised spacelike (Killing) vectors orthogonal to $\tilde k^\mu$: 
\be
l_{(i)}=\tilde l_{(i)}^\mu \partial_\mu =\partial_{z^{i}}.
\ee
Furthermore, we have that 
\begin{align}
\tilde{l}^\mu \partial_\mu\phi=0.
\end{align}

For the conformal frame metric, we now have a null Killing vector $\tilde k^\mu$ and covariantly constant spacelike vectors $\tilde l_{(i)}^\mu$. We can perform the GGV transform on conformal frame metric following \cite{GGV}, and use the technology of \cite{GGV} to compute the left and right hand sides of the transformed equations. Using this we discover  (i) the right set of transformation rules, (ii) conditions to be satisfied by the background spacetime in order for the technique to work, and (iii) the correct equation to be satisfied by the deforming scalar field. Once the technique is established in one frame, we can go to string or Einstein frame metric by appropriate  conformal transformation. 

Since conformal transformations play an important role in this discussion, let us recall the transformation rules for the covariant derivative and the Ricci tensor under a conformal transformation. 
For the conformal transformation $\widetilde{g}_{\mu\nu}=\Omega^2 g_{\mu\nu}$ in $n$-dimensions the covariant derivative transforms as
\be
\widetilde{\del}_\mu\omega_\nu =\del_\mu \omega_\nu - C^\rho_{\mu\nu}\omega_\rho, \label{cod_transform_conf}\\
\ee
where
\be
C^\rho_{\mu\nu}=\delta^\rho_\mu \del_\nu(\ln\Omega)+\delta^\rho_\nu \del_\mu(\ln\Omega)-g_{\mu\nu}g^{\rho\sigma} \del_\sigma(\ln\Omega), \label{connection_new}
\ee
and the Ricci tensor transforms as,
\bea
\nn 
\widetilde{R}_{\mu\nu}&=&R_{\mu\nu}-(n-2)\del_\mu\del_\nu(\ln\Omega)-g_{\mu\nu}g^{\rho\sigma}\del_\rho\del_\sigma(\ln\Omega)+(n-2)(\del_\mu \ln\Omega)(\del_\nu \ln\Omega)\label{Ricci_tranform_conf}\\
&& -(n-2)g_{\mu\nu}g^{\rho\sigma}(\del_\rho \ln\Omega)(\del_\sigma \ln\Omega).
\eea

Using these transformations we first obtain Einstein equations in the conformal frame: \be
\tilde{g}^{\rom{(C)}}_{\mu\nu}=e^{-\phi/2} g^{\rom{(E)}}_{\mu\nu}.\ee  
Einstein equations take the form
\be
\tilde{R}_{\mu\nu}=2\tilde{\del}_\mu \tilde{\del}_\nu \phi+\frac{1}{4} \tilde{g}^{\rho\alpha}\tilde{g}^{\sigma\beta} F_{\mu\rho\sigma} F_{\nu\alpha\beta}\label{conformal_frame_einstein_eqs}.
\ee

In the conformal frame, we postulate the generalised Garfinkle-Vachaspati (GGV),
\bea
\tilde g_{\mu\nu}&\to& \tilde g'_{\mu\nu} = \tilde g_{\mu\nu}+\Psi (\tilde k_{\mu}\tilde l_{\nu}+\tilde k_\nu \tilde l_\mu), \label{GGV_conf_1}\\
C_{\mu \nu} &\to& C'_{\mu \nu} =  C_{\mu \nu}-   \Psi \ (\tilde k_{\mu}\tilde l_{\nu}-\tilde k_{\nu}\tilde l_{\mu}).\label{GGV_conf_2}
\eea
We will see from the analysis below that the above transformation works when the background spacetime configuration satisfies a ``transversality'' condition:
\be
\tilde{k}^\mu \tilde{F}_{\mu\nu\rho}=-(d\tilde{k})_{\nu\rho}, \label{transversality}
\ee
and the scalar $\Psi$ satisfies,
\be
\tilde{\Box}\Psi+2(\partial_\mu\phi)\tilde{g}^{\mu\nu}(\partial_\nu\Psi)=0.
\ee

Two comments are in order. First, in the dilaton-free case, the scalar equation reduces to the minimally coupled massless scalar equation for $\Psi$. Second, remarkably, the transversality condition \eqref{transversality} in the conformal frame  is  the same as for the dilaton-free case considered in \cite{GGV}. We can check that the transversality condition is consistent with the Einstein equations -- it is the ``square root'' of a doubly contracted Einstein equations even in the present set-up. Contracting the left hand side of the Einstein equations \eqref{conformal_frame_einstein_eqs} twice with the null Killing vector $\tilde k^\mu$, we can simplify it to
\be
\tilde{ R}_{\mu\nu} \tilde{k}^\mu \tilde{k}^\nu=-\tilde{k}^\lambda\tilde{\Box} \tilde{k}_\lambda =\frac{1}{4} (\tilde{\del}^\rho \tilde{k}^\sigma-\tilde{\del}^\sigma \tilde{k}^\rho)(\tilde{\del}_\rho \tilde{k}_\sigma-\tilde{\del}_\sigma \tilde{k}_\rho).
\ee
Contracting the right hand side similarly, we first note that,
\be
\tilde{k}^\mu \tilde{k}^\nu \tilde{\del}_\mu \tilde{\del}_\nu \phi=0.
\ee
Therefore, the contracted Einstein equations simply reduce to
\be
(\tilde{\del}^\rho \tilde{k}^\sigma-\tilde{\del}^\sigma \tilde{k}^\rho)(\tilde{\del}_\rho \tilde{k}_\sigma-\tilde{\del}_\sigma \tilde{k}_\rho)= (\tilde{k}^\mu \tilde{F}_{\mu\rho\sigma}) (\tilde{k}^\nu{\tilde{F}_\nu}{}^{\rho\sigma}),
\ee
which is just the square of the transversality condition \eqref{transversality}.

To proceed further,  we need to compute the transformation of the Ricci tensor under the GGV. For this computation we simply follow the steps from appendix A.1 of  \cite{GGV} to find the transformed Ricci tensor in the conformal frame is\footnote{We have confirmed this transformation using Cadabra \cite{Cadabra1,Cadabra2} too.},
\begin{eqnarray}
\tilde{R}'_{\lambda\nu} 
 &=& \tilde{R}_{\lambda\nu}-\tilde{l}_\lambda[ \tilde{k}^\mu (\tilde{\del}_\nu\tilde{\del}_\mu \Psi)+\Psi \tilde{\Box} \tilde{k}_\nu ]-\tilde{l}_\nu [ \tilde{k}^\mu (\tilde{\del}_\lambda\tilde{\del}_\mu \Psi)+\Psi \tilde{\Box} \tilde{k}_\lambda ]\nonumber \\
&&+\frac{1}{2}(\tilde{\del}_\rho\Psi)(\tilde{\del}^\rho\Psi)\tilde{k}_\lambda \tilde{k}_\nu-\Psi^2(\tilde{\del}_\mu \tilde{k}^\rho)(\tilde{\del}_\rho \tilde{k}^\mu ) \tilde{l}_\lambda \tilde{l}_\nu-\frac{1}{2}\tilde{\Box}\Psi \tilde{S}_{\lambda\nu},
\label{transformed_ricci_tensor}
\end{eqnarray}
where we have introduced the notation
\bea
\tilde{S}_{\mu\nu} = \tilde{k}_\mu \tilde{l}_\nu + \tilde{k}_\nu \tilde{l}_\mu.
\eea
From this we can also check that Ricci scalar remains invariant under the GGV transform, i.e., $\tilde{R}'=\tilde{R}$.

 \subsection{Right hand side of Einstein equations}
Under the GGV  transform \eqref{GGV_conf_1}--\eqref{GGV_conf_2} dilaton remains invariant. The two-form field transforms as \eqref{GGV_conf_2}. For the convenience of notation we also put tildes on the two-form field in the conformal frame\footnote{Strictly speaking this notation is not needed, but since we need to raise and lower its indices it is a  useful notation.}
\be
\tilde{C} \to \tilde{C}'=\tilde{C}-\Psi \ \tilde{k}_\mu dx^\mu \wedge \tilde{l}_\nu dx^\nu. \label{GGV_conf_2_2}
\ee

To work out the transformation of the right hand side of the Einstein equation \eqref{conformal_frame_einstein_eqs},  we follow the steps from appendix A.2 of  \cite{GGV}. We introduce 
\be
\tilde{m}_{\mu\nu}=\tilde{k}_\mu \tilde{l}_\nu -\tilde{k}_\nu \tilde{l}_\mu,
 \ee
and write the transformed R-R two-form field \eqref{GGV_conf_2_2} as
\be
\tilde{C}'_{\mu \nu } =\tilde{C}_{\mu \nu}-\Psi \tilde{m}_{\mu \nu}.
\ee
We then find that
\be
\tilde{F}'_{\mu\nu\rho}
=\tilde{F}_{\mu\nu\rho}-\tilde{Q}_{\mu\nu\rho}-\Psi \tilde{P}_{\mu\nu\rho},
\ee
where
\bes
\bea
\tilde{Q}_{\mu\nu\rho}&=&(\partial_\mu\Psi)\tilde{m}_{\nu\rho}+(\partial_\rho\Psi)\tilde{m}_{\mu\nu}+(\partial_\nu\Psi)\tilde{m}_{\rho\mu }, \label{Q}\\
\tilde{P}_{\mu\nu\rho}&=&\partial_\mu \tilde{m}_{\nu\rho}+\partial_\rho \tilde{m}_{\mu\nu}+\partial_\nu \tilde{m}_{\rho\mu }. \label{P}
\eea
\ees
Now we need to raise the indices on $\tilde{F}_{\mu\nu\rho}$ and contract them appropriately with another $\tilde{F}_{\mu\nu\rho}$.  We find
\begin{eqnarray}
\tilde{F}'^{\sigma\eta\alpha}
&=&\tilde{F}^{\sigma\eta\alpha}-\tilde{Q}^{\sigma\eta\alpha},\label{F_upper}
\end{eqnarray}
and
\bea
\frac{1}{4}\tilde{F}'_{\lambda\alpha\beta}\tilde{F}'^{\delta\alpha\beta}&=&\frac{1}{4}\tilde{F}_{\lambda\alpha\beta}\tilde{F}^{\delta\alpha\beta}-[\tilde{l}^\delta (\tilde{\del}_\lambda\tilde{\del}_\beta\Psi) +\tilde{l}_\lambda(\tilde{\del}^\delta \tilde{\del}_\beta\Psi)]\tilde{k}^\beta \nn \\
&&+\frac{1}{2}(\tilde{\del}_\beta\Psi)(\tilde{\del}^\beta\Psi)\tilde{k}_\lambda \tilde{k}^\delta +\Psi \tilde{k}^\delta \tilde{k}^\alpha (\tilde{\del}_\lambda\tilde{\del}_\alpha\Psi)-\Psi \tilde{l}_\lambda \Box \tilde{k}^\delta.
\eea
From this last expression it can be easily seen that 
\be
\tilde{F}'_{\alpha\beta\gamma}\tilde{F}'^{\alpha\beta\gamma} = \tilde{F}_{\alpha\beta\gamma}\tilde{F}^{\alpha\beta\gamma}.
\ee
The transformation of the dilaton term in \eqref{conformal_frame_einstein_eqs} is similarly computed:
\be
2{\tilde\del}'_\mu {\tilde\del}'_\nu \phi =2{\tilde\del}_\mu{\tilde\del}_\nu\phi+{\tilde g}^{\rho\sigma}({\tilde \del}_\sigma \Psi) {\tilde S}_{\mu\nu }(\partial_\rho\phi)-2\Psi [({\tilde \del}_\mu {\tilde k}^\rho) {\tilde l}_\nu+({\tilde \del}_\nu {\tilde k}^\rho) {\tilde l}_\mu]({\tilde \del}_\rho\phi).
\ee

Comparing with the left hand side \eqref{transformed_ricci_tensor}, we see that the variations of both sides match provided,
\be
-\frac{1}{2}{\tilde S}_{\mu\nu}{\tilde \Box}\Psi ={\tilde g}^{\rho\sigma}({\tilde \del}_\sigma \Psi) {\tilde S}_{\mu\nu }(\partial_\rho\phi)-2\Psi [({\tilde \del}_\mu {\tilde k}^\rho) {\tilde l}_\nu+({\tilde \del}_\nu {\tilde k}^\rho) {\tilde l}_\mu]({\tilde \del}_\rho\phi). \label{tensor_eq}
\ee
This looks like a non-trivial tensor equation for the scalar $\Psi$. Fortunately, making use of the background Einstein equations this can be simplified. The term $2(\tilde \del_\mu \tilde k^\rho)(\tilde \del_\rho\phi)$ is simplified to 
\be
2(\tilde \del_\mu \tilde k^\rho)(\tilde \del_\rho\phi)=-2\tilde k^\rho(\tilde \del_\mu\tilde \del_\rho\phi). \label{term_2}
\ee
Next, using the background Einstein equation we can write the right hand side of equation \eqref{term_2} as,
\bea
-2\tilde k^\rho(\tilde \del_\mu\tilde \del_\rho\phi)=-2\tilde k^\rho\left(\tilde R_{\mu\rho}-\frac{1}{4}\tilde F_\mu {}^{\alpha\beta}\tilde F_{\rho\alpha\beta}\right)=-2\tilde k^\rho \tilde R_{\mu\rho}+\frac{1}{2}\tilde F_\mu {}^{\alpha\beta}(-\tilde n_{\alpha\beta})
\eea
where
\be
\tilde{n}_{\mu\nu}=\tilde{\del}_\mu \tilde{k}_\nu-\tilde{\del}_\nu \tilde{k}_\mu 
\ee
Using the identities
\be
\tilde k^\rho \tilde R_{\mu\rho}=-\tilde \Box \tilde k_\mu, \qquad \qquad \tilde n_{\alpha\beta} \tilde{F}_\mu{}^{\alpha\beta}=4\tilde \Box \tilde k_\mu ,
\ee
we get,
\be
-2 \tilde k^\rho(\tilde \del_\mu\tilde \del_\rho\phi)=0.
\ee
Similarly,
\be
2(\tilde \del_\nu \tilde k^\rho)(\tilde \del_\rho\phi)=0.
\ee
As a result of these simplifications, tensor equation \eqref{tensor_eq} becomes a scalar equation,
\be
\tilde{\Box}\Psi+2(\partial_\mu\phi)\tilde{g}^{\mu\nu}(\partial_\nu\Psi)=0\label{scalar_eqn},
\ee
equivalently,
\be
\tilde{\del}_\mu(e^{2\phi}\tilde{g}^{\mu\nu}(\partial_\nu\Psi))=0\label{scalar_eqn_conf}.
\ee

\subsection{Matter field equations}
   
The 2-form field equation in Einstein frame is \eqref{eq_matter}. Under conformal transformation from Einstein frame to the conformal frame we note that 
\be
\tilde{\del}_\mu\omega^{\mu\nu\rho} =\del_\mu \omega^{\mu\nu\rho}+C^\mu_{\mu\gamma}\omega^{\gamma\nu\rho}+C^\nu_{\mu\alpha}\omega^{\mu\alpha\rho}+C^\rho_{\mu\beta}\omega^{\mu\nu\beta},
\ee
with $C^\nu_{\mu\alpha}$ given in equation \eqref{connection_new}. A simple calculation using these equations gives that the 2-form field equation in the conformal frame can be written as,
\be
\tilde{\del}_\mu(e^{2\phi} \tilde{F}^{\mu\nu\rho})=0. \label{eq_matter_conf}
\ee

To check the validity of our generalised GV transform, we need to check that the matter field equations transforms covariantly. 
Under the generalised GV transform the dilaton remains invariant while the 3-form field strength with all the three contravariant (upper) indices transforms as \eqref{F_upper}
\be
\tilde{F}'^{\mu\nu\rho}=\tilde{F}^{\mu\nu\rho}-\tilde{Q}^{\mu\nu\rho}.
\ee
Therefore, under the GGV transformation we have the following transformation of the left hand side of equation \eqref{eq_matter_conf} (more details can be found in appendix A.3 of \cite{GGV}),
\be
\tilde{\del}'_\mu ( e^{2\phi}\tilde{F}'^{\mu\nu\rho})
=\tilde{\del}_\mu(e^{2\phi} (\tilde{F}^{\mu\nu\rho}-\tilde{Q}^{\mu\nu\rho}))\label{matter_simp},
\ee
where 
\be
\tilde{Q}^{\mu\nu\rho}
=(\tilde{\del}^\mu\Psi)\tilde{m}^{\nu\rho}+(\tilde{\del}^\nu\Psi)\tilde{m}^{\rho\mu}+(\tilde{\del}^\rho\Psi)\tilde{m}^{\mu\nu},
\ee
and recall that $\tilde{m}^{\mu\nu}=\tilde{k}^\mu \tilde{l}^\nu -\tilde{k}^\nu \tilde{l}^\mu =m^{\mu\nu}$. Expanding the above expression and using \eqref{scalar_eqn} one can show that 
\be
\tilde{\del}'_\mu ( e^{2\phi}\tilde{F}'^{\mu\nu\rho})
= \tilde{\del}_\mu ( e^{2\phi}\tilde{F}^{\mu\nu\rho}) =0.
\ee

\subsection{Summary in different R-R and NS-NS frames}
\label{app:summary}
Now that we have established that the GGV transform is a valid solution generating technique, we can express it in any frame we like. 

\paragraph{Conformal frame:}
The generalised Garfinkle-Vachaspati transform is
\bea
\tilde g_{\mu\nu}&\to& \tilde g'_{\mu\nu} = \tilde g_{\mu\nu}+\Psi (\tilde k_{\mu}\tilde l_{\nu}+\tilde k_\nu \tilde l_\mu), \\
C_{\mu \nu} &\to& C'_{\mu \nu} =  C_{\mu \nu}-   \Psi \ (\tilde k_{\mu}\tilde l_{\nu}-\tilde k_{\nu}\tilde l_{\mu}).
\eea
with scalar $\Psi$ satisfying 
\be
\tilde{\del}_\mu(e^{2\phi}\tilde{g}^{\mu\nu}(\partial_\nu\Psi))=0,
\ee
and the 2-form C-field satisfying the transversality condition
\be
\tilde{k}^\mu \tilde{F}_{\mu\nu\rho}=-(d\tilde{k})_{\nu\rho}.
\ee
In this frame the vector $\tilde l^\mu$ is covariantly constant and unit normalised.

   \paragraph{Einstein frame:}

By performing the conformal transformation by the factor $g^{\rom{(E)}}_{\mu\nu}=e^{\phi/2}\tilde{g}_{\mu\nu}$ we obtain the GGV in the Einstein frame. In this frame the spacelike vector $l^\mu$ is not covariantly constant.
It satisfies
\be
\del^\rom{(E)}_\mu l^\rom{(E)}_\nu=\frac{1}{4}[l^\rom{(E)}_\nu (\partial_\mu \phi)-l^\rom{(E)}_\mu (\partial_\nu \phi)].
\ee
It of course satisfies the Killing equation,
\be
\del^\rom{(E)}_\mu l^\rom{(E)}_\nu +\del^\rom{(E)}_\nu l^\rom{(E)}_\mu =0,
\ee
and is normalised as $l_\rom{(E)}^\mu l^\rom{(E)}_\mu =e^{\phi/2}$.

Under the conformal transformation from the conformal frame to Einstein frame the scalar $\Psi$ equation simply becomes
\be
\Box^\rom{(E)}\Psi=0.
\ee
The transversality condition \eqref{transversality} becomes
\be
k_\rom{(E)}^\mu F_{\mu\nu\rho}=-d(e^{-\phi/2}k^\rom{(E)})_{\nu\rho}.
\ee
The GGV transform takes the form
 \bea
 g_{\mu\nu}^\rom{(E)}&\to& g_{\mu\nu}^\rom{(E)}+\Psi e^{-\phi/2}(k^\rom{(E)}_\mu l^\rom{(E)}_\nu+k^\rom{(E)}_\nu l^\rom{(E)}_\mu ),\\
C&\to& C-\Psi e^{-\phi} (k^\rom{(E)}_\mu l^\rom{(E)}_\nu-k^\rom{(E)}_\nu l^\rom{(E)}_\mu ).
\eea

\paragraph{String frame:}

By performing the conformal transformation by the factor $g^{\rom{(S)}}_{\mu\nu}=e^{\phi/2}g^{\rom{(E)}}_{\mu\nu}$ we obtain the GGV transform in the string frame. In this case we present some more details as this is the set-up we  exclusively work with in the main text of the paper. In this frame too, the spacelike vector $l_{\rom{(S)}}^\mu$ is not covariantly constant.
It satisfies
\be
\del^\rom{(S)}_\mu l^\rom{(S)}_\nu=\frac{1}{2}[l^\rom{(S)}_\nu (\partial_\mu \phi)-l^\rom{(S)}_\mu (\partial_\nu \phi)],
\ee
\be
\del^\rom{(S)}_\mu l^\rom{(S)}_\nu +\del^\rom{(S)}_\nu l^\rom{(S)}_\mu =0,
\ee
It is now normalised as $l_\rom{(S)}^\mu l^\rom{(S)}_\mu =e^{\phi}$.

Under the conformal transformation from  Einstein frame to string frame the scalar $\Psi$ equation becomes
\be
\Box^{\rom{(S)}}\Psi-2(\partial_\mu\phi)g_{\rom{(S)}}^{\mu\nu}(\partial_\nu\Psi)=0,
\ee
equivalently
\be
\del^\rom{(S)}_{\mu} (e^{-2\phi}g_\rom{(S)}^{\mu\nu} \partial_\nu\Psi)=0.
\ee
The transversality condition \eqref{transversality} becomes
\be
k_\rom{(S)}^\mu F_{\mu\nu\rho}=-d(e^{-\phi}k^\rom{(S)})_{\nu\rho}.
\ee
The GGV transform takes the form
 \bea
 g_{\mu\nu}^\rom{(S)}&\to& g_{\mu\nu}^\rom{(S)}+\Psi e^{-\phi}(k^\rom{(S)}_\mu l^\rom{(S)}_\nu+k^\rom{(S)}_\nu l^\rom{(S)}_\mu ),\\
C&\to& C-\Psi e^{-2\phi} (k^\rom{(S)}_\mu l^\rom{(S)}_\nu-k^\rom{(S)}_\nu l^\rom{(S)}_\mu ).
\eea

For ease of reference we write the string frame equations of motion (and omit the superscript $\rom{(S)}$ for convenience). The IIB string frame action with RR 2-form $C^{(2)}$ with $F^{(3)} = d C^{(2)}$ is 
\be
S=\frac{1}{16 \pi G}\int d^{10}x\sqrt{-g}\left[e^{-2\Phi}[R+4(d\Phi)^2]-\frac{1}{12}F_{\mu \nu \rho}F^{\mu \nu \rho}\right],
\ee
and the resulting equations of motion are
\bea
R_{\mu \nu} + 2 \nabla_\mu \nabla_\nu \Phi &=& \frac{1}{4} e^{2\Phi}\left(F_{\mu  \rho \sigma} F_{\nu}{}^{ \rho \sigma} - \frac{1}{6} F_{\rho \sigma \kappa}F^{\rho \sigma \kappa} g_{\mu \nu}  \right), \\
\nabla_\mu F^{\mu \nu \rho} &= &0, \\
R + 4 \nabla^2 \Phi - 4 (\nabla \Phi)^2 &=& 0.
\eea

\paragraph{NS-NS sector string frame:}

Embedding of interest of the six-dimensional theory \eqref{action_6d}
 in the ten-dimensional NS-NS sector string frame is as follows
\be
ds^2_{\rom{(S)}} = e^{-\phi} ds^2_6 + ds_4^2, \label{embedding_NS_string}
\ee
with ten-dimensional dilaton,
\be
\Phi = -\phi.
\ee
The six-dimensional 2-form field is now the 2-form B-field with zero components in the four torus directions.

In this embedding the torus Killing vectors are unit normalised and are covariantly constant.  In this set-up the GGV transform takes the form,
\bea
g_{\mu\nu}&\to& g_{\mu\nu}+\Psi (k_\mu l_\nu + k_\nu l_\mu )\\
B_{\mu\nu} &\to& B_{\mu\nu}-\Psi (k_\mu l_\nu - k_\nu l_\mu )
\eea
The transversality condition reads,
\be
k^\mu  H_{\mu\nu\rho}=-(dk)_{\nu\rho},
\ee
and the scalar wave equation for the field $\Psi$ becomes
\be
\del_\mu (e^{-2\phi}g^{\mu\nu}\del_\nu\Psi)=0.
\ee

\paragraph{NS-NS sector Einstein frame:}
In Einstein frame, embedding \eqref{embedding_NS_string} reads,
\be
ds^2_{\rom{(E)}} = e^{-\Phi/2} ds^2_{\rom{(S)}}  =e^{\phi/2} ds^2_{\rom{(S)}}  = e^{-\phi/2} ds^2_6 +e^{\phi/2} ds_4^2. \label{embedding_NS_einstein}
\ee
We note that this metric is same as \eqref{metric_einstein_10d}. The two embeddings are related by S-duality:
\begin{align}
g^\rom{(E)}_{\mu\nu}&\to g^\rom{(E)}_{\mu\nu},& \Phi &\to -\Phi,& C_{\mu\nu}&\to B_{\mu\nu}.
\end{align}
 In this set-up the GGV transform takes the form,
\bea
g^\rom{(E)}_{\mu\nu}&\to& g^\rom{(E)}_{\mu\nu}+\Psi e^{\phi/2}(k^\rom{(E)}_\mu l^\rom{(E)}_\nu + k^\rom{(E)}_\nu l^\rom{(E)}_\mu),\\
B_{\mu\nu}&\to& B_{\mu\nu}-\Psi e^{\phi} (k^\rom{(E)}_\mu l^\rom{(E)}_\nu - k^\rom{(E)}_\nu l^\rom{(E)}_\mu).
\eea
The transversality condition reads,
$k_\rom{(E)}^\mu  H_{\mu\nu\rho}=-(d(e^{\phi/2}k^\rom{(E)}))_{\nu\rho}, $ and the scalar wave equation for the field $\Psi$ becomes $\Box^{\rom{(E)}}\Psi=0. $


\begin{thebibliography}{99}
   
   
\bibitem{Sen:1995in} 
  A.~Sen,
  ``Extremal black holes and elementary string states,''
  Mod.\ Phys.\ Lett.\ A {\bf 10}, 2081 (1995)
  doi:10.1142/S0217732395002234
  [hep-th/9504147].
  
  
  
\bibitem{Strominger:1996sh} 
  A.~Strominger and C.~Vafa,
  ``Microscopic origin of the Bekenstein-Hawking entropy,''
  Phys.\ Lett.\ B {\bf 379}, 99 (1996)
  doi:10.1016/0370-2693(96)00345-0
  [hep-th/9601029].


    
\bibitem{Sen:2007qy} 
  A.~Sen,
  ``Black Hole Entropy Function, Attractors and Precision Counting of Microstates,''
  Gen.\ Rel.\ Grav.\  {\bf 40}, 2249 (2008)
  doi:10.1007/s10714-008-0626-4
  [arXiv:0708.1270 [hep-th]].
  
  

  
\bibitem{Mandal:2010cj} 
  I.~Mandal and A.~Sen,
  ``Black Hole Microstate Counting and its Macroscopic Counterpart,''
  Nucl.\ Phys.\ Proc.\ Suppl.\  {\bf 216}, 147 (2011)
  [Class.\ Quant.\ Grav.\  {\bf 27}, 214003 (2010)]
  doi:10.1088/0264-9381/27/21/214003
  [arXiv:1008.3801 [hep-th]].
  
\bibitem{Dabholkar:2012zz} 
  A.~Dabholkar and S.~Nampuri,
  ``Quantum black holes,''
  Lect.\ Notes Phys.\  {\bf 851}, 165 (2012)
  doi:10.1007/978-3-642-25947-05
  [arXiv:1208.4814 [hep-th]].


  
\bibitem{Banerjee:2009uk} 
  N.~Banerjee, I.~Mandal and A.~Sen,
  ``Black Hole Hair Removal,''
  JHEP {\bf 0907}, 091 (2009)
  doi:10.1088/1126-6708/2009/07/091
  [arXiv:0901.0359 [hep-th]].

\bibitem{Jatkar:2009yd} 
  D.~P.~Jatkar, A.~Sen and Y.~K.~Srivastava,
  ``Black Hole Hair Removal: Non-linear Analysis,''
  JHEP {\bf 1002}, 038 (2010)
  doi:10.1007/JHEP02(2010)038
  [arXiv:0907.0593 [hep-th]].

\bibitem{Mathur:2018tib} 
  S.~D.~Mathur and D.~Turton,
  ``The fuzzball nature of two-charge black hole microstates,''
  arXiv:1811.09647 [hep-th].
  
    
\bibitem{David:2006yn} 
  J.~R.~David and A.~Sen,
  ``CHL Dyons and Statistical Entropy Function from D1-D5 System,''
  JHEP {\bf 0611}, 072 (2006)
  doi:10.1088/1126-6708/2006/11/072
  [hep-th/0605210].
  


\bibitem{Garfinkle:1990jq} 
  D.~Garfinkle and T.~Vachaspati,
  ``Cosmic string traveling waves,''
  Phys.\ Rev.\ D {\bf 42}, 1960 (1990).
  doi:10.1103/PhysRevD.42.1960
  
\bibitem{Kaloper:1996hr} 
  N.~Kaloper, R.~C.~Myers and H.~Roussel,
  ``Wavy strings: Black or bright?,''
  Phys.\ Rev.\ D {\bf 55}, 7625 (1997)
  doi:10.1103/PhysRevD.55.7625
  [hep-th/9612248].
  
  


\bibitem{Dabholkar:1995nc} 
  A.~Dabholkar, J.~P.~Gauntlett, J.~A.~Harvey and D.~Waldram,
  ``Strings as solitons and black holes as strings,''
  Nucl.\ Phys.\ B {\bf 474}, 85 (1996)
  doi:10.1016/0550-3213(96)00266-0
  [hep-th/9511053].



\bibitem{Horowitz:1996th} 
  G.~T.~Horowitz and D.~Marolf,
  ``Counting states of black strings with traveling waves,''
  Phys.\ Rev.\ D {\bf 55}, 835 (1997)
  doi:10.1103/PhysRevD.55.835
  [hep-th/9605224].


  
\bibitem{Banados:1999tw} 
  M.~Banados, A.~Chamblin and G.~W.~Gibbons,
  ``Branes, AdS gravitons and Virasoro symmetry,''
  Phys.\ Rev.\ D {\bf 61}, 081901 (2000)
  doi:10.1103/PhysRevD.61.081901
  [hep-th/9911101].
  
\bibitem{Hubeny:2003ug} 
  V.~E.~Hubeny and M.~Rangamani,
  ``Horizons and plane waves: A Review,''
  Mod.\ Phys.\ Lett.\ A {\bf 18}, 2699 (2003)
  doi:10.1142/S0217732303012428
  [hep-th/0311053].
  
    
\bibitem{Balasubramanian:2010ys} 
  V.~Balasubramanian, J.~Parsons and S.~F.~Ross,
  ``States of a chiral 2d CFT,''
  Class.\ Quant.\ Grav.\  {\bf 28}, 045004 (2011)
  doi:10.1088/0264-9381/28/4/045004
  [arXiv:1011.1803 [hep-th]].
  

 
\bibitem{Mathur:2005zp} 
  S.~D.~Mathur,
  ``The Fuzzball proposal for black holes: An Elementary review,''
  Fortsch.\ Phys.\  {\bf 53}, 793 (2005)
  doi:10.1002/prop.200410203
  [hep-th/0502050].
  
  
\bibitem{Bena:2007kg} 
  I.~Bena and N.~P.~Warner,
  ``Black holes, black rings and their microstates,''
  Lect.\ Notes Phys.\  {\bf 755}, 1 (2008)
    doi:10.1007/978-3-540-79523-0-1
  [hep-th/0701216].


\bibitem{Skenderis:2008qn} 
  K.~Skenderis and M.~Taylor,
  ``The fuzzball proposal for black holes,''
  Phys.\ Rept.\  {\bf 467}, 117 (2008)
  doi:10.1016/j.physrep.2008.08.001
  [arXiv:0804.0552 [hep-th]].
  
  
\bibitem{Chowdhury:2010ct} 
  B.~D.~Chowdhury and A.~Virmani,
  ``Modave Lectures on Fuzzballs and Emission from the D1-D5 System,''
  arXiv:1001.1444 [hep-th].


\bibitem{Callan:1995hn} 
  C.~G.~Callan, J.~M.~Maldacena and A.~W.~Peet,
  ``Extremal black holes as fundamental strings,''
  Nucl.\ Phys.\ B {\bf 475}, 645 (1996)
  doi:10.1016/0550-3213(96)00315-X
  [hep-th/9510134].
 
   
\bibitem{Lunin:2001jy} 
  O.~Lunin and S.~D.~Mathur,
  ``AdS / CFT duality and the black hole information paradox,''
  Nucl.\ Phys.\ B {\bf 623}, 342 (2002)
  doi:10.1016/S0550-3213(01)00620-4
  [hep-th/0109154].
  
\bibitem{Mathur:2011gz} 
  S.~D.~Mathur and D.~Turton,
  ``Microstates at the boundary of AdS,''
  JHEP {\bf 1205}, 014 (2012)
  doi:10.1007/JHEP05(2012)014
  [arXiv:1112.6413 [hep-th]].
  
  
  \bibitem{Mathur:2012tj} 
  S.~D.~Mathur and D.~Turton,
  ``Momentum-carrying waves on D1-D5 microstate geometries,''
  Nucl.\ Phys.\ B {\bf 862}, 764 (2012)
  doi:10.1016/j.nuclphysb.2012.05.014
  [arXiv:1202.6421 [hep-th]].
  
  
\bibitem{Lunin:2012gp} 
  O.~Lunin, S.~D.~Mathur and D.~Turton,
  ``Adding momentum to supersymmetric geometries,''
  Nucl.\ Phys.\ B {\bf 868}, 383 (2013)
  doi:10.1016/j.nuclphysb.2012.11.017
  [arXiv:1208.1770 [hep-th]].
   
\bibitem{GGV} 
  D.~Mishra, Y.~K.~Srivastava and A.~Virmani,
  ``A Generalised Garfinkle-Vachaspati Transform,''
  Gen.\ Rel.\ Grav.\  {\bf 50}, no. 12, 155 (2018)
  doi:10.1007/s10714-018-2477-y
  [arXiv:1808.04981 [hep-th]]. In topical collection ``The Fuzzball Paradigm'' Editors: S.~Mathur, D.~Tutron, A.~Virmani.

\bibitem{Duff} 
  M.~J.~Duff, J.~T.~Liu and J.~Rahmfeld,
``Four-dimensional string-string-string triality,''
  Nucl.\ Phys.\ B {\bf 459}, 125 (1996)
  doi:10.1016/0550-3213(95)00555-2
  [hep-th/9508094].

\bibitem{Pol1} 
  J.~Polchinski,
  ``String theory. Vol. 1: An introduction to the bosonic string,''
  doi:10.1017/CBO9780511816079
  
\bibitem{Pol2} 
  J.~Polchinski,
  ``String theory. Vol. 2: Superstring theory and beyond,''
  doi:10.1017/CBO9780511618123

 
\bibitem{Cadabra1} 
  K.~Peeters,
  ``Introducing Cadabra: A Symbolic computer algebra system for field theory problems,''
  hep-th/0701238.
  
  
\bibitem{Cadabra2} 
  K.~Peeters, ``Cadabra2: computer algebra for field theory revisited,'' Journal of Open Source Software, {\bf 3 (32)}, 1118, (2018) https://doi.org/10.21105/joss.01118
  
  
    
  
\bibitem{gms1} 
  S.~Giusto, S.~D.~Mathur and A.~Saxena,
  ``Dual geometries for a set of 3-charge microstates,''
  Nucl.\ Phys.\ B {\bf 701}, 357 (2004)
  doi:10.1016/j.nuclphysb.2004.09.001
  [hep-th/0405017].
  
  
\bibitem{gms2} 
  S.~Giusto, S.~D.~Mathur and A.~Saxena,
  ``3-charge geometries and their CFT duals,''
  Nucl.\ Phys.\ B {\bf 710}, 425 (2005)
  doi:10.1016/j.nuclphysb.2005.01.009
  [hep-th/0406103].



 
 
\bibitem{GMR} 
  J.~B.~Gutowski, D.~Martelli and H.~S.~Reall,
  ``All Supersymmetric solutions of minimal supergravity in six- dimensions,''
  Class.\ Quant.\ Grav.\  {\bf 20}, 5049 (2003)
  doi:10.1088/0264-9381/20/23/008
  [hep-th/0306235].
  
  
\bibitem{Lam:2018jln} 
  H.~Het Lam and S.~Vandoren,
  ``BPS solutions of six-dimensional (1, 0) supergravity coupled to tensor multiplets,''
  JHEP {\bf 1806}, 021 (2018)
  doi:10.1007/JHEP06(2018)021
  [arXiv:1804.04681 [hep-th]].
  
  
\bibitem{Cano:2018wnq} 
  P.~A.~Cano and T.~Ort\'in,
  ``All the supersymmetric solutions of ungauged $\mathcal{N} = (1,0),d=6$ supergravity,''
  arXiv:1804.04945 [hep-th].
  
  
    
  
\bibitem{Horowitz:1994ei} 
  G.~T.~Horowitz and A.~A.~Tseytlin,
  ``On exact solutions and singularities in string theory,''
  Phys.\ Rev.\ D {\bf 50}, 5204 (1994)
  doi:10.1103/PhysRevD.50.5204
  [hep-th/9406067].

\bibitem{Horowitz:1994rf} 
  G.~T.~Horowitz and A.~A.~Tseytlin,
  ``A New class of exact solutions in string theory,''
  Phys.\ Rev.\ D {\bf 51}, 2896 (1995)
  doi:10.1103/PhysRevD.51.2896
  [hep-th/9409021].
   
 \bibitem{Lunin:2002iz} 
  O.~Lunin, J.~M.~Maldacena and L.~Maoz,
  ``Gravity solutions for the D1-D5 system with angular momentum,''
  hep-th/0212210.
  
  \bibitem{Kanitscheider:2007wq} 
  I.~Kanitscheider, K.~Skenderis and M.~Taylor,
  ``Fuzzballs with internal excitations,''
  JHEP {\bf 0706}, 056 (2007)
  doi:10.1088/1126-6708/2007/06/056
  [arXiv:0704.0690 [hep-th]].
  
\bibitem{Bena:2005ay} 
  I.~Bena and P.~Kraus,
  ``Microstates of the D1-D5-KK system,''
  Phys.\ Rev.\ D {\bf 72}, 025007 (2005)
  doi:10.1103/PhysRevD.72.025007
  [hep-th/0503053].
 
\bibitem{Virmani:2012kw} 
  A.~Virmani,
  ``Subtracted Geometry From Harrison Transformations,''
  JHEP {\bf 1207}, 086 (2012)
  doi:10.1007/JHEP07(2012)086
  [arXiv:1203.5088 [hep-th]].
  
\bibitem{Bena:2017geu} 
  I.~Bena, E.~Martinec, D.~Turton and N.~P.~Warner,
  ``M-theory Superstrata and the MSW String,''
  JHEP {\bf 1706}, 137 (2017)
  doi:10.1007/JHEP06(2017)137
  [arXiv:1703.10171 [hep-th]].
\end{thebibliography}
\end{document}